# The specific heat of superfluids near the transition temperature

Norbert Schultka and Efstratios Manousakis

*Department of Physics and Center for Materials Research and Technology*
*Florida State University, Tallahassee, Florida 32306*

February 15, 1995

**Abstract**

The specific heat of the $x-y$ model is studied on cubic lattices of sizes $L \times L \times L$ and on lattices $L \times L \times H$ with $L \gg H$ (i.e. on lattices representing a film geometry) using the Cluster Monte Carlo method. Periodic boundary conditions were applied in all directions. In the cubic case we obtained the ratio of the critical exponents $\alpha/\nu$ from the size dependence of the energy density at the critical temperature $T_\lambda$. Using finite–size scaling theory, we find that while for both geometries our results scale to universal functions, these functions differ for the different geometries. We compare our findings to experimental results and results of renormalization group calculations.

## 1 Introduction

Physical systems which exhibit a second order phase transition and are confined in a finite geometry (e.g. a cubic or film geometry) are thought to be well described by the finite–size scaling theory at temperatures close to the critical temperature $T_\lambda$ [1]. This scaling theory states that finite–size effects can be observed when the bulk correlation length $\xi$ becomes of the order of the finite extent $L$ of the system, e.g. in a cubic geometry the length of the edges of the cube plays the role of $L$. For a physical observable $O$ this intuitive assumption can be cast into the following formula [2]:

$$\frac{O(t,L)}{O(t,L=\infty)} = f\left(\frac{L}{\xi(t,L=\infty)}\right), \qquad (1)$$

where $\xi(t,L=\infty)$ is the correlation length for the infinite size system, $t$ is the reduced temperature, and $f$ is a universal function. For example, the most singular behavior of the correlation length close to the critical temperature is given by $\xi(t) = \xi_0^{\pm}|t|^{-\nu}$. In this case using Eq. (1) with $O(t,L) = \xi(t,L)$ we obtain:

$$\xi(t,L) = |t|^{-\nu} f_\xi(|t|L^{1/\nu}), \qquad (2)$$

where the prefactor $\xi_0^{\pm}$ has been absorbed in the definition of the universal function $f_\xi(x)$.

A physical system, which has been widely used to experimentally test the finite–size scaling theory, is liquid $^4He$ since the superfluid density $\rho_s$ and the specific heat $c$ can be measured to a very high accuracy. However, the experimental verification of the finite–size scaling theory is somewhat controversial. Rhee, Gasparini, and Bishop measured the superfluid density of very thick helium films [3] and showed that the data did not follow the form (1) using a value of $\nu$ reasonably close to the expected value $\nu = 0.67$. Similarly, in early measurements of the specific heat of $^4He$ in finite geometries very different critical exponents from the expected values were found [4] indicating that phenomenological finite-size scaling cannot be applied in a straightforward way. In order to clarify the situation renormalization group calculations for the standard Landau–Ginzburg functional in different geometries with Dirichlet boundary conditions have been undertaken [5, 6, 7, 8]. New specific heat measurements [9] and also a reanalysis [10] of the old specific heat data [4] show good agreement between the renormalization group calculations reported in [5, 6, 7] and those data. Furthermore new experiments on liquid $^4He$ confined in a film geometry under microgravity conditions are planned [11] in order to determine the finite–size scaling of the specific heat. Thus, it is desirable to obtain reliable results for the specific heat in finite–size helium systems by means of numerical investigations.

In this paper we perform a numerical study of the scaling behavior of the specific heat of $^4He$ in a cubic ($L \times L \times L$ size lattices) and in a film geometry ($L \times L \times H$ size lattices with $L >> H$) at temperatures close to the critical temperature $T_\lambda$. Since $^4He$ belongs to the universality class of the $x-y$ model [12], we use this model to compute the specific heat at temperatures near $T_\lambda$ using the 1–cluster Monte Carlo method [13]. The $x-y$ model on a lattice is defined as

$$\mathcal{H} = -J \sum_{\langle i,j \rangle} \vec{s}_i \cdot \vec{s}_j, \qquad (3)$$

where the summation is over all nearest neighbors, $\vec{s} = (\cos\theta, \sin\theta)$ is a two–component vector constrained to the unit circle, and $J$ sets the energy scale.

The critical exponents of the three–dimensional $x-y$ model have been determined by high–temperature expansions [14] and Monte Carlo simulations [15, 16, 17, 18]. The importance of vortex lines for the phase transition was investigated in Ref. [19]. A renormalization group approach based on vortex lines [20] derives the critical properties of the three–dimensional $x-y$ model from the interaction of vortex lines. The anisotropic three–dimensional $x-y$ model ($J_x = J_y \neq J_z$) has also been studied [21], and a crossover from three–dimensional to two–dimensional behavior was found with respect to the ratio $J_z/J_x$. The Villain model, which is in the same universality class as the $x-y$ model, has been studied in a film geometry where the correlation length in the disordered phase was used to extract the thickness–dependent critical temperature [22]. The authors of Ref. [23] computed the universal scaling function of the superfluid density of helium confined in a film geometry using the $x-y$ model and examined the crossover properties from three to two dimensions of the superfluid density.

In this work we compute the specific heat $c(T, L)$ of the $x-y$ model on various cubic lattices $L \times L \times L$. We deduce its critical exponent $\alpha$ from the size–dependence of the energy density at the critical temperature, estimate the bulk value $c(T_\lambda, \infty)$, and check the scaling hypothesis for the specific heat with respect to $L$. We compare the resulting universal function to recent renormalization group calculations [26]. The specific heat for a film geometry, i.e. for various $L^2 \times H$ lattices with $L \to \infty$, is also computed and the scaling behavior of the specific heat with respect to $H$ is studied. We compare the universal scaling function for the film geometry directly to the experimental results of Refs. [9].

The article is organized as follows. In the next section we introduce the definition of the specific heat and the energy density and briefly discuss the Monte Carlo method. Section 3 discusses the finite–size scaling properties of the specific heat. In section 4 we deduce the critical exponents and check the scaling assumption of the specific heat for the cubic geometry. Section 5 is devoted to the film geometry, and the last section summarizes our results.

## 2 Definition of the physical quantities and Monte Carlo method

We define the energy density of our model as follows:

$$E = \langle e \rangle = 3 - \frac{1}{V} \left\langle \sum_{\langle i,j \rangle} \vec{s}_i \cdot \vec{s}_j \right\rangle , \tag{4}$$

where $V = L^3$ for the cubes and $V = HL^2$ for the film geometry. The specific heat can be written as ($k_B = 1$):

$$c = V\beta^2(\langle e^2 \rangle - \langle e \rangle^2), \tag{5}$$

where $\beta = J/(k_B T)$.

The thermal averages, denoted by the angular brackets, are computed according to

$$\langle O \rangle = Z^{-1} \int \prod_i d\theta_i \, O[\theta] \exp\left(-\frac{\mathcal{H}}{k_B T}\right). \tag{6}$$

$O[\theta]$ denotes the dependence of the physical observable $O$ on the configuration $\{\theta_i\}$, and the partition function $Z$ is given by

$$Z = \int \prod_i d\theta_i \, \exp\left(-\frac{\mathcal{H}}{k_B T}\right). \tag{7}$$

The multi–dimensional integrals in the expressions (6) and (7) are computed with Wolff's 1–cluster algorithm Monte Carlo method [13]. We computed the specific heat at various temperatures on $L^3$–lattices for $L = 20, 30, 40$ and on $L^2 \times H$–lattices for $L = 40, 60, 100$ and $H = 6, 8, 10$. At the critical temperature the energy density and the specific heat were computed on $L^3$ lattices for $L = 10, 15, 20, 25, 30, 35, 40, 45, 50, 60, 80$. Periodic boundary conditions were applied in all directions. We carried out of the order of 20,000 thermalization steps and of the order of 500,000 measurements. The calculations were performed on a heterogeneous environment of computers including Sun, IBM RS/6000 and DEC alpha AXP workstations and a Cray–YMP.

## 3 Finite–size scaling properties of the specific heat

Let us first consider the $x - y$ model in a cube whose edges are of length $L$. In such a geometry the specific heat starts feeling the finite size of the cube when the bulk correlation length $\xi$ becomes comparable to the length $L$. At temperatures close to the critical temperature $T_\lambda$ and for $L \to \infty$ we can write [2]

$$\frac{c(t,L)}{c(t,\infty)} = G(tL^{1/\nu}), \tag{8}$$

where the reduced temperature $t = T/T_\lambda - 1$, $G$ is a universal function, and $\nu$ is the critical exponent of the correlation length. In the limit $L \to \infty$ and at a fixed value of $t$ we obtain $G(\pm\infty) = 1$. Now we leave $L$ fixed but carry out the limit $t \to \pm 0$ assuming $c(t,\infty) \propto |t|^{-\alpha}$ with $\alpha > 0$. We obtain

$$\lim_{t\to\pm 0} c(t,L) \propto \lim_{t\to\pm 0} |t|^{-\alpha} G(tL^{1/\nu}), \tag{9}$$

and the fact that $c(0,L)$ is finite at $t = 0$ implies that

$$\lim_{x\to\pm 0} G(x) \propto |x|^{\alpha}, \tag{10}$$

thus

$$c(0, L) \propto L^{\alpha/\nu}. \tag{11}$$

Experiments on superfluid $^4He$ [24, 25] indicate that $\nu = 0.6705$, and via the hyperscaling relation $\alpha = 2 - 3\nu = -0.0115 < 0$, thus $c(0, \infty)$ is finite. In order to write $c(t, L)$ in a scaling form similar to (8), notice:

$$c(t, \infty) - c(0, \infty) \propto |t|^{-\alpha}, \tag{12}$$

with $\alpha < 0$. Since the scaling theory deals only with the most "singular" terms of a physical quantity when the critical point is approached, the following scaling form suggests itself:

$$\frac{c(t, L) - c(0, \infty)}{c(t, \infty) - c(0, \infty)} = G(tL^{1/\nu}) \tag{13}$$

or

$$c(t, L) = c(0, \infty) + |t|^{-\alpha} g(tL^{1/\nu}). \tag{14}$$

Keeping $t$ fixed, we find $\lim_{x \to \pm\infty} g(x) = \tilde{c}_1^{\pm}$, which is finite. For fixed $L$ and $t \to \pm 0$ we obtain the behavior of Eq. (10) and

$$c(0, L) = c(0, \infty) + c_1 L^{\alpha/\nu}. \tag{15}$$

After defining the scaling function $\tilde{g}(x) = |x|^{-\alpha} g(x)$ we obtain

$$c(t, L) = c(0, \infty) + L^{\alpha/\nu} \tilde{g}(tL^{1/\nu}). \tag{16}$$

This enables us to reexpress the scaling form (13) as follows:

$$\frac{c(t, L) - c(0, \infty)}{c(0, L) - c(0, \infty)} = G_L(tL^{1/\nu}). \tag{17}$$

Note that $G_L(0) = 1$.

Let us finally derive the relationship between the scaling function $G_L(tL^{1/\nu})$ and the scaling function $f_1(tL^{1/\nu})$ used in Refs. [5, 6]. This function is defined as follows:

$$c(t, L) - c(t_0, \infty) = L^{\alpha/\nu} f_1(tL^{1/\nu}), \tag{18}$$

where $t_0 = (\xi_0^+/L)^{1/\nu}$, i.e. $t_0$ is the reduced temperature where the correlation length is equal to the system size $L$. Using Eqs. (15) and (17) we can write

$$c(t, L) - c(0, \infty) = c_1 L^{\alpha/\nu} G_L(tL^{1/\nu}). \tag{19}$$

In order to make Eq. (19) consistent with Eq. (10), we have to require that if we keep $t$ fixed and take the limit $L \to \infty$

$$\lim_{x \to \pm\infty} G_L(x) = g_\infty^{\pm} |x|^{-\alpha}, \tag{20}$$

where $g_\infty^{\pm}$ is a constant. Thus, at $L = \infty$ Eq. (19) can be written as:

$$c(t, \infty) - c(0, \infty) = \tilde{c}_1^{\pm} |t|^{-\alpha}, \tag{21}$$

where $\tilde{c}_1^{\pm} = c_1 g_\infty^{\pm}$. Evaluating expression (21) at $t_0$, solving for $c(0, \infty)$ and inserting the result into Eq. (19) yields

$$c(t, L) - c(t_0, \infty) = L^{\alpha/\nu} c_1 \left\{ G_L(tL^{1/\nu}) - g_\infty^{+} (\xi_0^+)^{-\alpha/\nu} \right\}. \tag{22}$$

Comparing this expression and the definition for $f_1(x)$ given by Eq. (18) we find

$$f_1(x) = c_1 \left\{ G_L(x) - g_\infty^{+} (\xi_0^+)^{-\alpha/\nu} \right\}. \tag{23}$$

These scaling forms are valid for the film confining geometry also. In the case of a film geometry we need to replace $L$ by $H$ in the scaling equations (13), (17), and (23) because the relevant scale is $H$ when $L \gg H$.

# 4 The cubic geometry

In this section we investigate the finite–size scaling behavior of the specific heat of the $x-y$ model in the cubic geometry $L^3$.

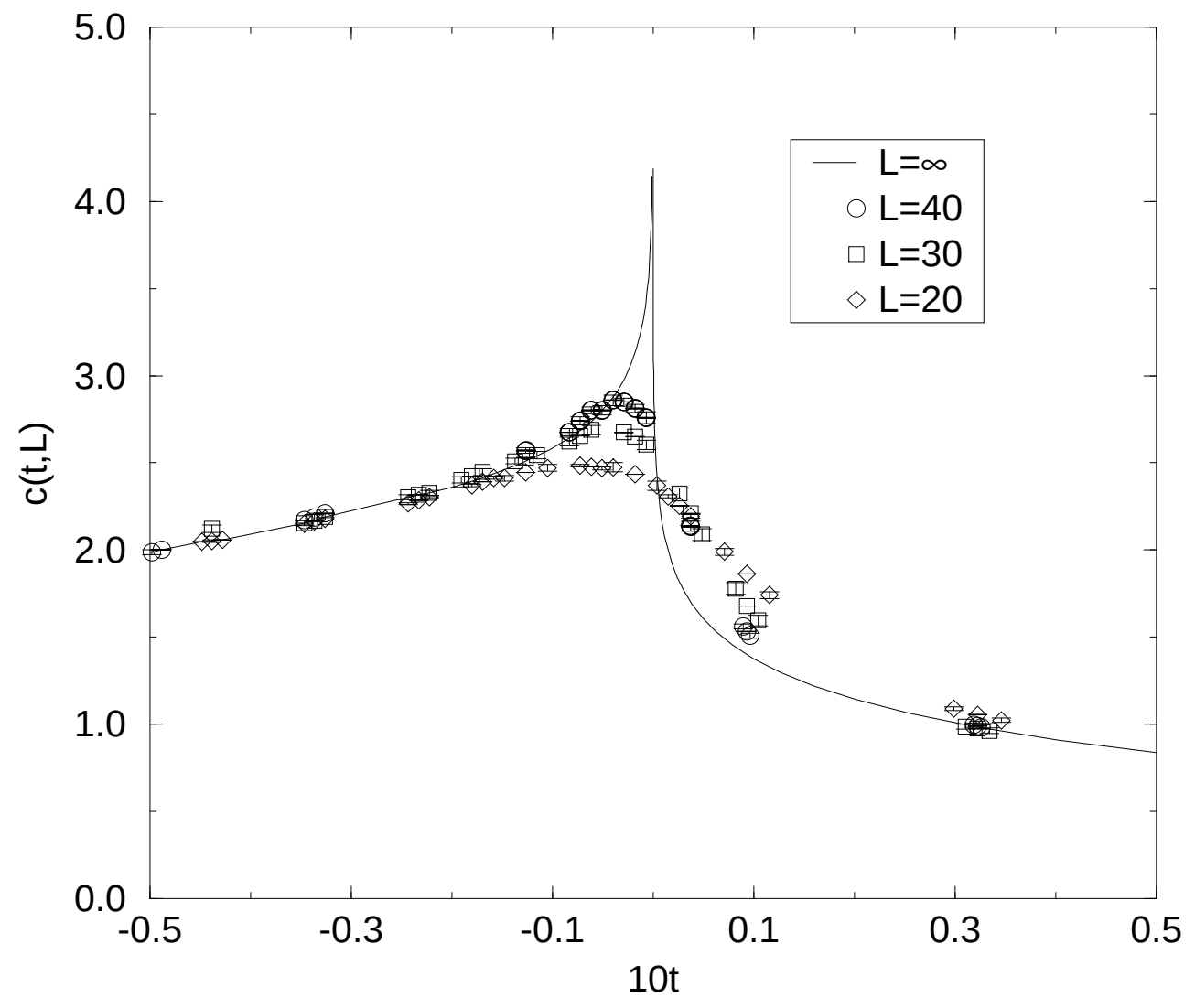

Figure 1: The specific heat for various size lattices $L$ as a function of the reduced temperature. The solid lines represent $L=\infty$ deduced from our Monte Carlo calculation. $T_\lambda = 2.2017$.

In Fig.1 we show our data for the specific heat. As a comparison we also plotted the bulk behavior of the specific heat (solid line). The steps leading to this curve are given below in this section.

| $L$ | $c(0,L)$ | $E(0,L)$ |
|---|---|---|
| 10 | 1.9454(69) | 1.95435(32) |
| 15 | 2.180(11) | 1.98078(22) |
| 20 | 2.362(10) | 1.99176(18) |
| 25 | 2.465(13) | 1.99709(21) |
| 30 | 2.579(18) | 2.00059(13) |
| 35 | 2.621(19) | 2.00257(15) |
| 40 | 2.754(30) | 2.00464(28) |
| 45 | 2.789(34) | 2.00522(13) |
| 50 | 2.783(31) | 2.00618(10) |
| 60 | 2.967(45) | 2.00712(10) |
| 80 | 3.048(43) | 2.00882(11) |

Table 1: The Monte Carlo results for the specific heat $c(0,L)$ and the energy density $E(0,L)$ at the critical temperature $T_\lambda = 2.2017$.

In order to find the universal function $G_L(x)$ we need to know $c(0,\infty)$. This quantity can be found by calculating the specific heat at the critical temperature for various lattice sizes $L$ and fitting the data to the form (15). We take $T_\lambda = 2.2017$ as the critical temperature [18]. Table 1 contains the values for $c(0, L)$ and for $E(0, L)$. Since the specific heat is very sensitive to fluctuations, it has a relatively large error, making it very difficult to extract the very small exponent $\alpha/\nu$. Therefore we decided to use another procedure to find the values of the critical exponents. A quantity which is closely related to the specific heat is the energy density $E$. It is advantageous to use the energy density data because of the small error bars involved in its calculation. These error bars are two orders of magnitude smaller than the error bars of the specific heat. From the expression

$$c(t, L) = \frac{\partial E(t, L)}{\partial T}, \tag{24}$$

we obtain by integrating (16) up to a constant

$$E(t, L) = c(0, \infty)T + L^{(\alpha-1)/\nu} T_\lambda D(tL^{1/\nu}), \tag{25}$$

where $dD(x)/dx = \tilde{g}(x)$. For $t \to 0$ we obtain

$$E(0, L) = E_0 + E_1 L^{(\alpha-1)/\nu}. \tag{26}$$

| data points | $E_0$ | $E_1$ | $1/\nu$ | $\alpha/\nu$ | $\chi^2$ | $Q$ |
|---|---|---|---|---|---|---|
| 10 | 2.0111(2) | -1.80(15) | 1.479(40) | -0.029(20) | 1.58 | 0.15 |
| 9 | 2.0110(4) | -1.81(38) | 1.487(81) | -0.0258(75) | 1.76 | 0.12 |
| 8 | 2.0111(7) | -1.80(10) | 1.48(25) | -0.029(55) | 1.92 | 0.10 |
| 7 | 2.011(2) | -1.8(37) | 1.48(76) | -0.029(88) | 2.56 | 0.05 |

Table 2: Fitted values of the parameters entering expression (26). $Q$ is the goodness of the fit.

| data points | $c(0, \infty)$ | $c_1$ | $\chi^2$ | $Q$ |
|---|---|---|---|---|
| 10 | 21.33(50) | -20.57(54) | 1.51 | 0.15 |
| 9 | 20.45(66) | -19.61(72) | 1.12 | 0.35 |
| 8 | 20.72(94) | -19.9(10) | 1.28 | 0.26 |
| 7 | 20.2(13) | -19.3(15) | 1.47 | 0.20 |

Table 3: Fitted values of the parameters entering expression (15). $\alpha/\nu = -0.0258$. $Q$ is the goodness of the fit.

The results of the fits of the energy density data to the expression (26) are given in Table 2. In the fits we have subsequently excluded values of the energy density corresponding to smaller and smaller lattices. Because of the size of the error bars of the ratio $\alpha/\nu$ and $1/\nu$ we cannot make a definite statement that $\alpha/\nu < 0$ and $1/\nu < 1.5$. The fitting parameters, however, become stable when the energy density data obtained for lattice sizes $L \geq 20$ (9 data points) are used for the fits. Despite the large error bars, we always find $\alpha/\nu < 0$ and $1/\nu < 1.5$. The parameters for the fit including 9 data points and shown in Fig.2 are

$$\begin{aligned} E_0 &= 2.0110 \pm 0.0004, \\ E_1 &= -1.81 \pm 0.38, \\ 1/\nu &= 1.487 \pm 0.081, \qquad (27) \\ \alpha/\nu &= -0.0258 \pm 0.0075. \qquad (28) \end{aligned}$$

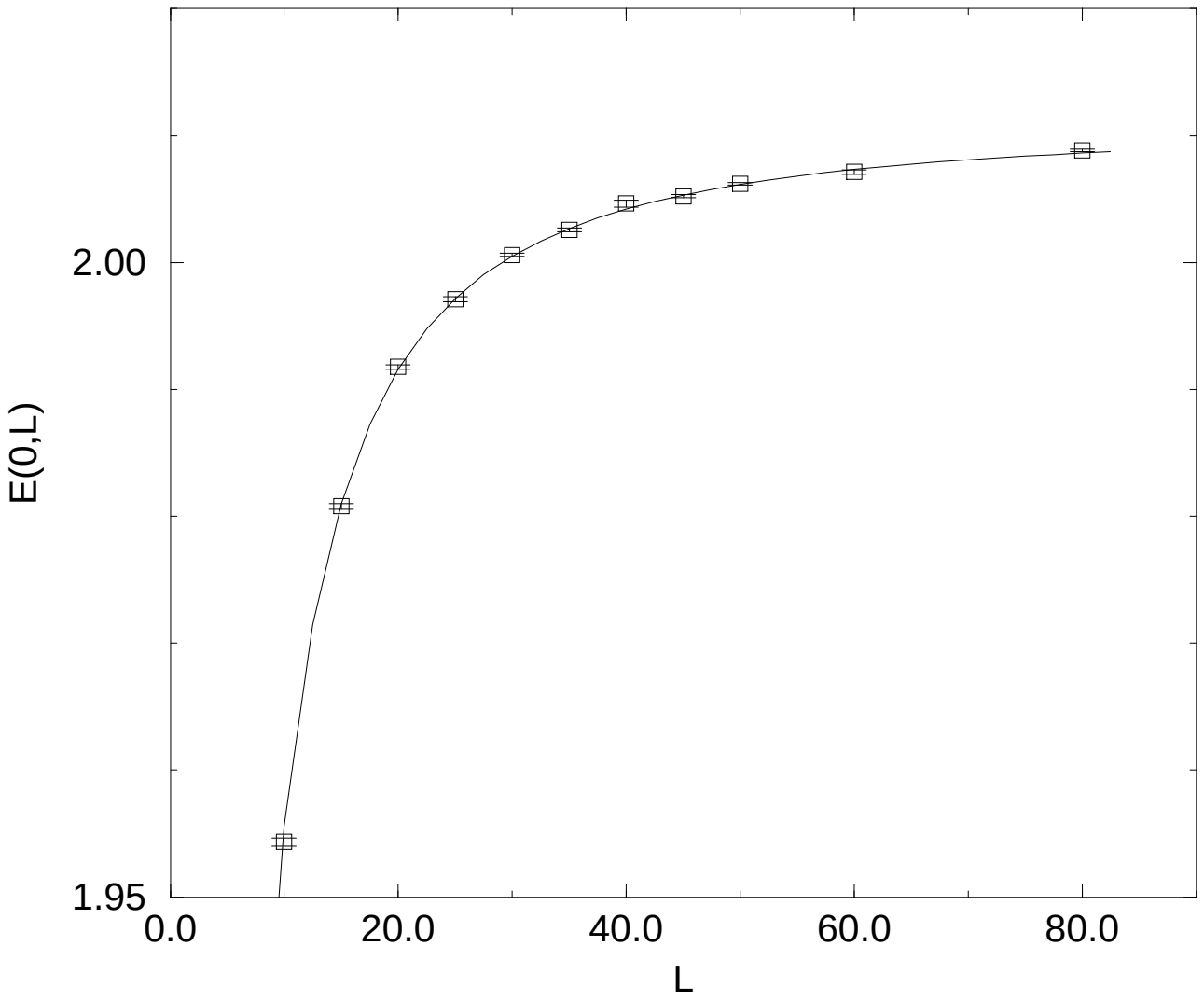

Figure 2: The energy density $E(0, L)$ at the critical temperature $T_\lambda = 2.2017$ as a function of $L$. The solid curve represents the fit to (26) (9 data points included).

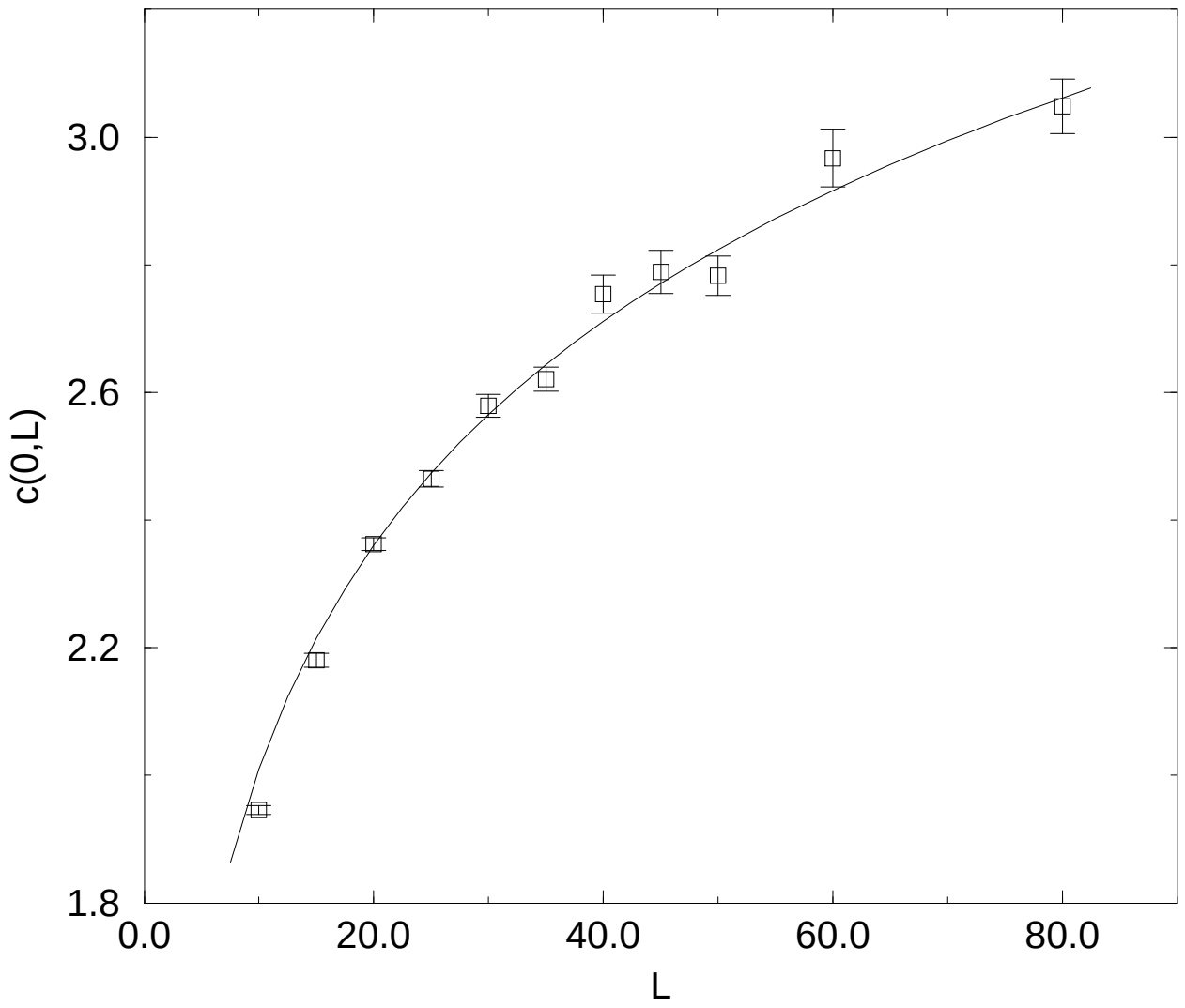

Figure 3: The specific heat $c(0, L)$ at the critical temperature $T_\lambda = 2.2017$ as a function of $L$. The solid curve represents the fit to (15) (9 data points included). $\alpha/\nu = -0.0258$.

Within error bars the hyperscaling assumption is fulfilled. The experimental value for $\alpha/\nu$ is $\alpha/\nu = -0.0172$ [24], and an earlier experiment gave $\alpha/\nu = -0.0225$ [25]. Having determined $\alpha/\nu$, we can turn to fitting the specific heat data to the expression (15). We fixed the value of $\alpha/\nu$ to the previously determined value $\alpha/\nu = -0.0258$. Table 3 contains the fitting results. If we exclude the data corresponding to the two smallest lattices we obtain

$$c(0,\infty) = 20.45 \pm 0.66, \tag{29}$$
$$c_1 = -19.61 \pm 0.72, \tag{30}$$

and the fit is shown in Fig.3.

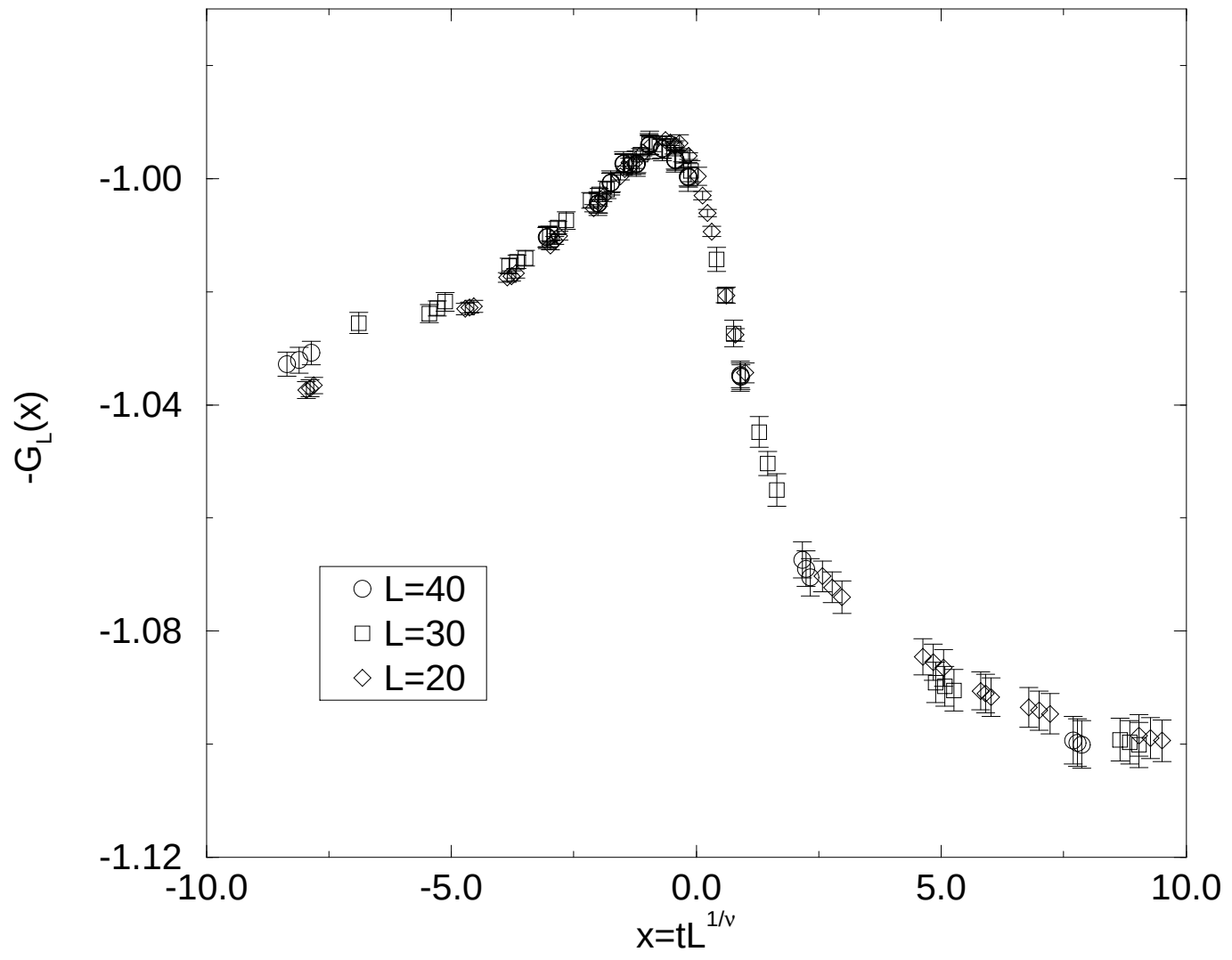

Figure 4: The scaling function $-G_L(x)$ (cf. Eq. (17)) for the cubic geometry. $c(0,\infty) = 20.45$, $1/\nu = 1.487$, $T_\lambda = 2.2017$, the values for $c(0,L)$ are taken from Table 1.

In order to check the validity of the scaling form (17) we plot $(c(t,L) - c(0,\infty))/(c(0,\infty) - c(0,L))$ versus $tL^{1/\nu}$ for different lattice sizes $L^3$ in Fig.4. We used the values for $c(0,L)$ given in Table 1 and $c(0,\infty)$ as determined above (Eq. (29)). As expected the data points for the three lattices $20^3$, $30^3$, and $40^3$ collapse onto one universal curve $G_L(tL^{1/\nu})$ in the range $-10 < tL^{1/\nu} < 10$.

It is interesting to repeat the fits described above using the experimentally determined critical exponents $\nu = 0.6705$ and $\alpha = -0.0115$ [24]. The result of the fit of the specific heat data corresponding to lattices of size $L \geq 20$ to the expression (15) is

$$c(0,\infty) = 30.3 \pm 1.0, \tag{31}$$
$$c_1 = -29.4 \pm 1.1. \tag{32}$$
$$\tag{33}$$

Fig.5 shows the scaling plot where $\nu = 0.6705$ and $c(0,\infty) = 30.3$. The data of the specific heat for the $20^3$, $30^3$, and $40^3$ lattice collapse onto one universal curve. We see, that the value of $c(0,\infty)$ is strongly effected by the value of the ratio $\alpha/\nu$. A much wider range of lattice sizes is necessary in order to determine these values more accurately. However, the scaling property, i.e. the data collapse onto one universal curve, is rather insensitive to the precise value of the ratio $\alpha/\nu$ and thus to the precise value of $c(0,\infty)$. This is demonstrated in Figs.4 and 5. Thus, for our lattice sizes the data collapse cannot be used to determine the value of the critical exponents more accurately.

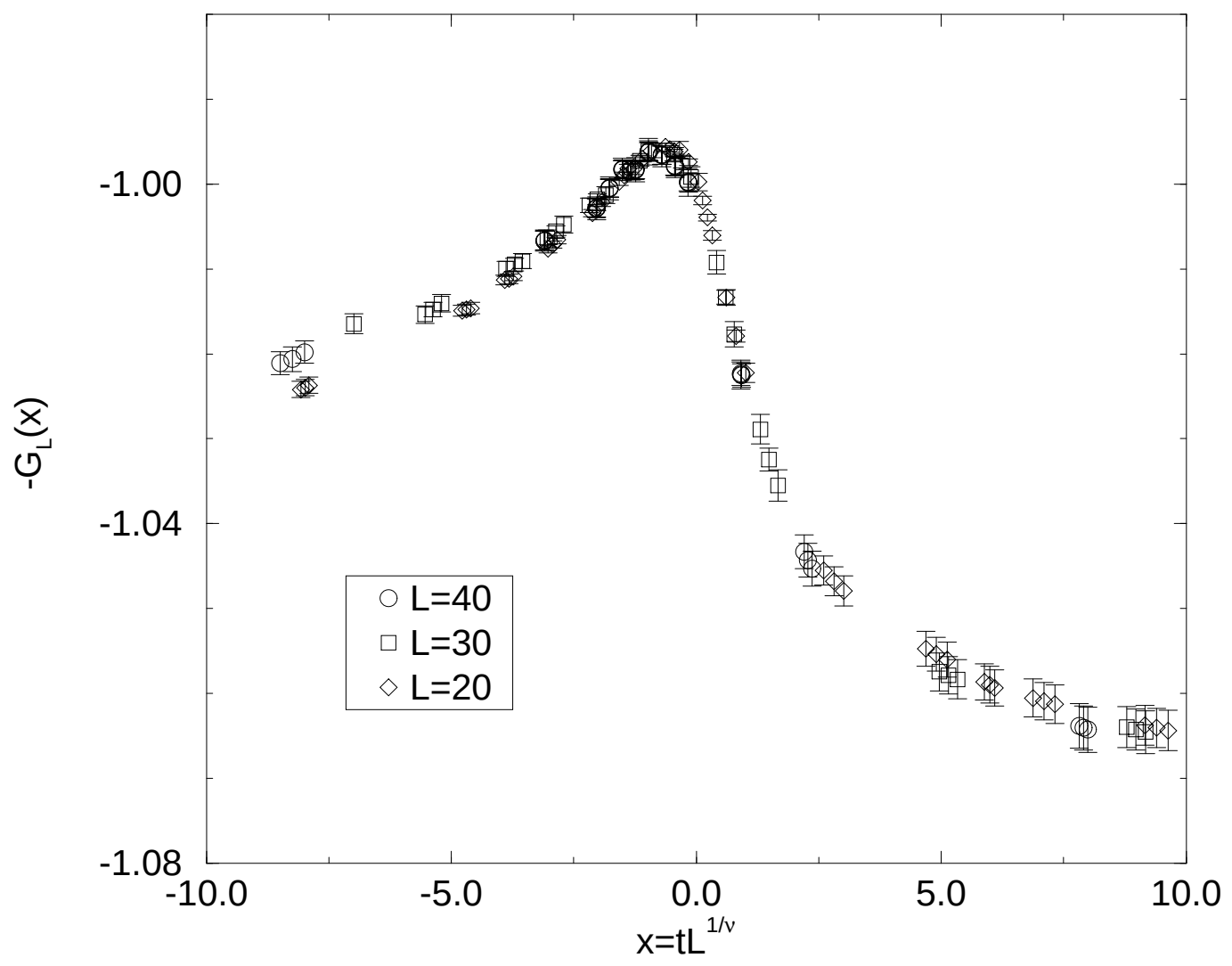

Figure 5: The scaling function $-G_L(x)$ (cf. Eq. (17)) for the cubic geometry. $c(0,\infty) = 30.3$, $\nu = 0.6705$, $T_\lambda = 2.2017$, the values for $c(0, L)$ are taken from Table 1.

In Fig.6 we compare the results of the renormalization group calculations of Ref. [26] (solid line) to the results of our Monte Carlo simulation for the lattice with $L = 48$. We computed the values for the specific heat on this lattice from our scaling function $G_L(x)$ (cf. Fig.5). The agreement between the renormalization group calculations and the Monte Carlo results is satisfactory.

We now compute the function $f_1(x)$ defined by Eq. (23) for the $x - y$ model. This function can be obtained from Eq. (23) with $\nu = 0.6705$, $c_1 = -29.4$, and $\xi_0^+ = 0.498$ from Ref. [16]. We estimate the value for $g_\infty^+$ as follows. Eq. (20) implies that if we plot $x^\alpha G_L(x)$ versus $x$ for large enough positive values of $x$ the function $x^\alpha G_L(x)$ should approach the finite value $g_\infty^+$. We obtain $g_\infty^+ = 1.0378(5)$ and thus

$$f_1(x) = -29.4\,[G_L(x) - 1.025]\,. \tag{34}$$

This function is shown in Fig.7. The solid line in Fig.7 represents the result of the renormalization group calculation of Ref. [26] (the scaling function $P_c(x)$ given there is related to $f_1(x)$ by the relation $f_1(x) = P_c(x) + 19.7$). Also here the agreement between the renormalization group calculation and the Monte Carlo results is satisfactory. We would like to note that the shape of the function $f_1(x)$ is rather insensitive to the precise value of $\alpha/\nu$ and thus to the precise value of $c_1$ (as was the scaling property of the specific heat), i.e. the scaling function $G_L(x)$ given in Fig.4 yields almost the same function $f_1(x)$ as shown in Fig.7. Therefore the scaling function $f_1(x)$ cannot be used to determine the critical exponents more accurately. However, the function $f_1(x)$ lends itself well to compare experimental results to the results of our calculation because in both cases it is hard to determine $\alpha$ and thus $c_1$ to a better accuracy.

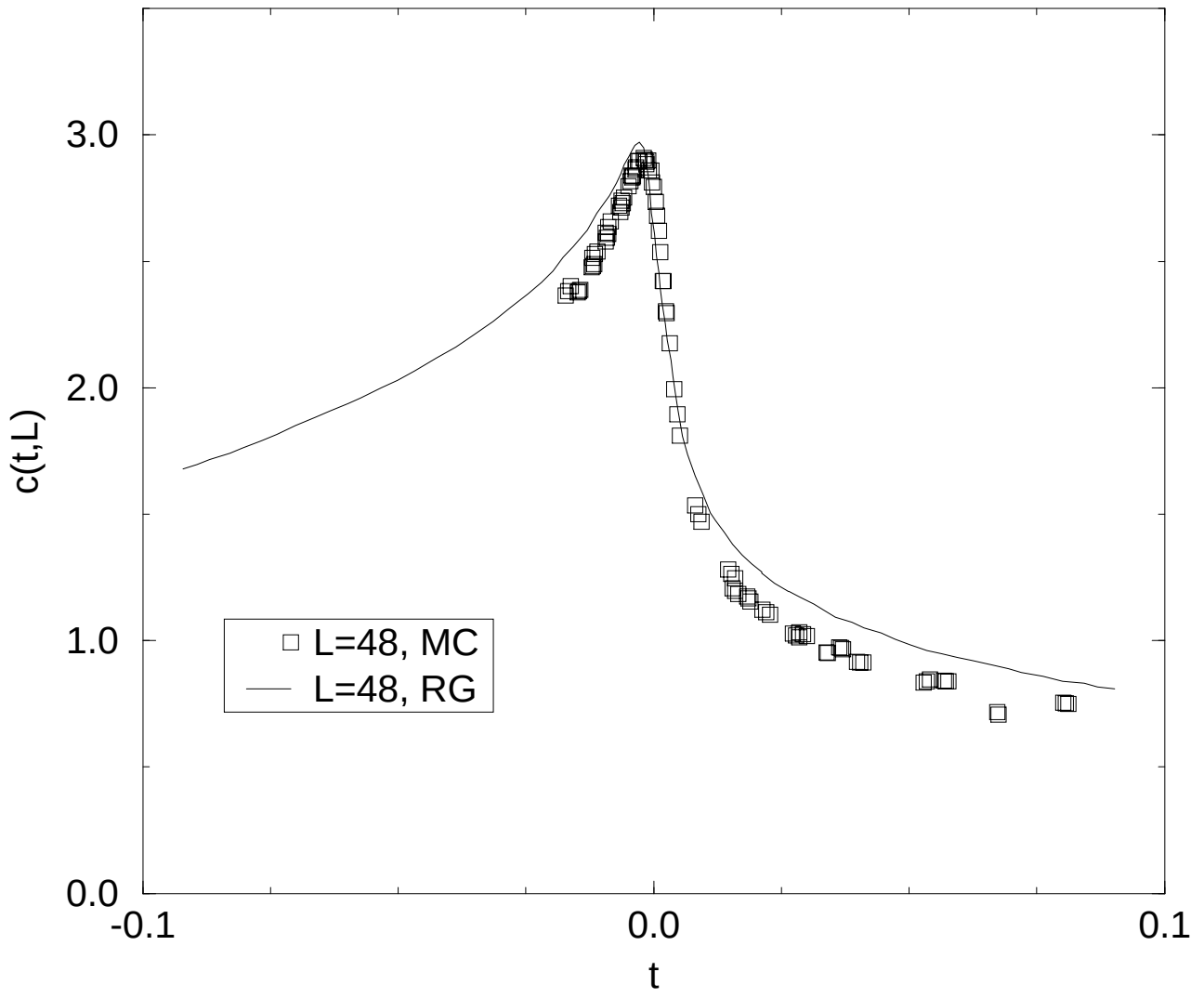

Figure 6: Comparison of the results of this work (MC) and the renormalization group (RG) calculations of Ref. [26] (solid line) for the specific on a $L = 48$ lattice.

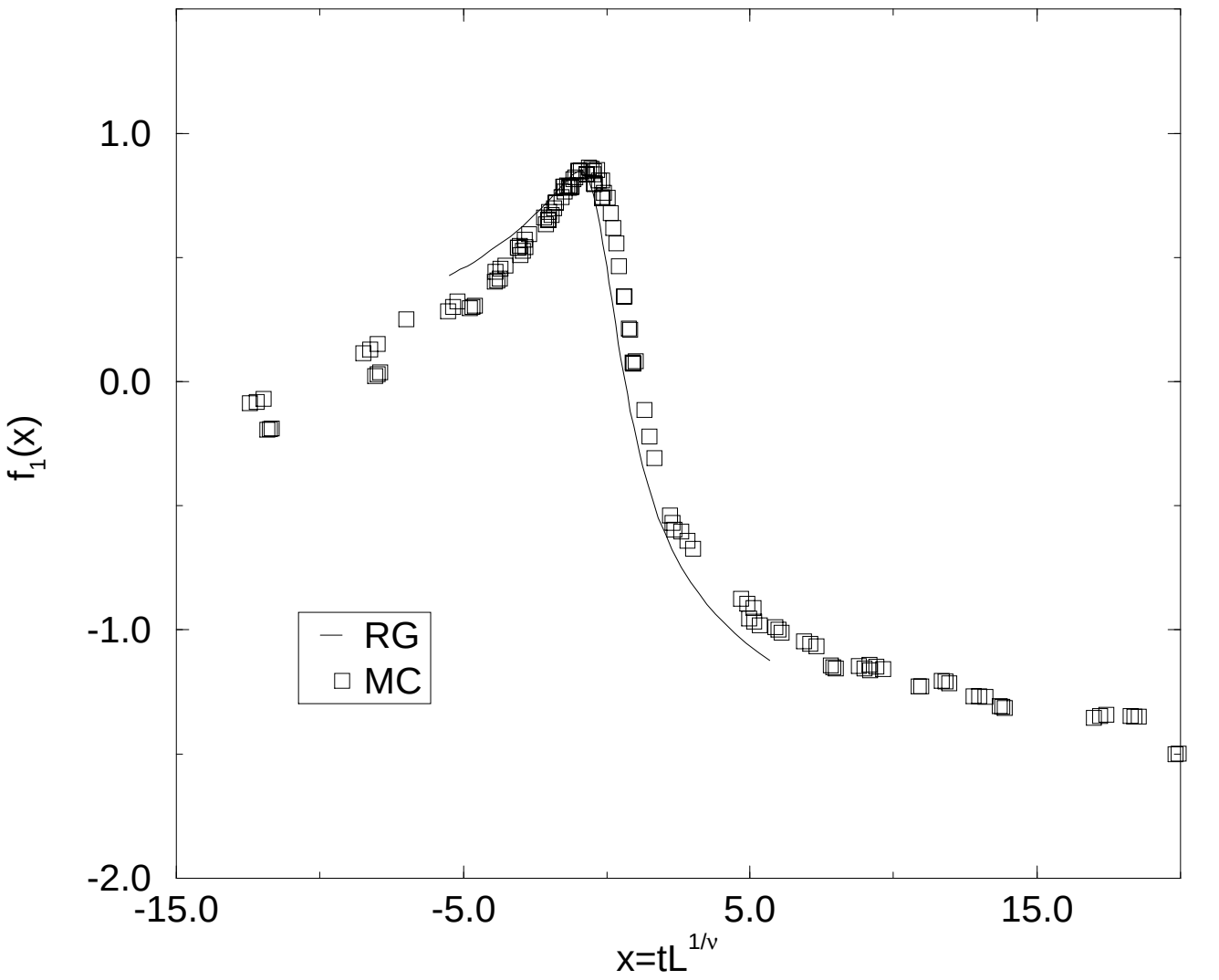

Figure 7: The function $f_1(x)$ for the cubic geometry. The solid line is the result of the renormalization group (RG) calculation of Ref. [26], the squares represent our Monte Carlo (MC) simulation.

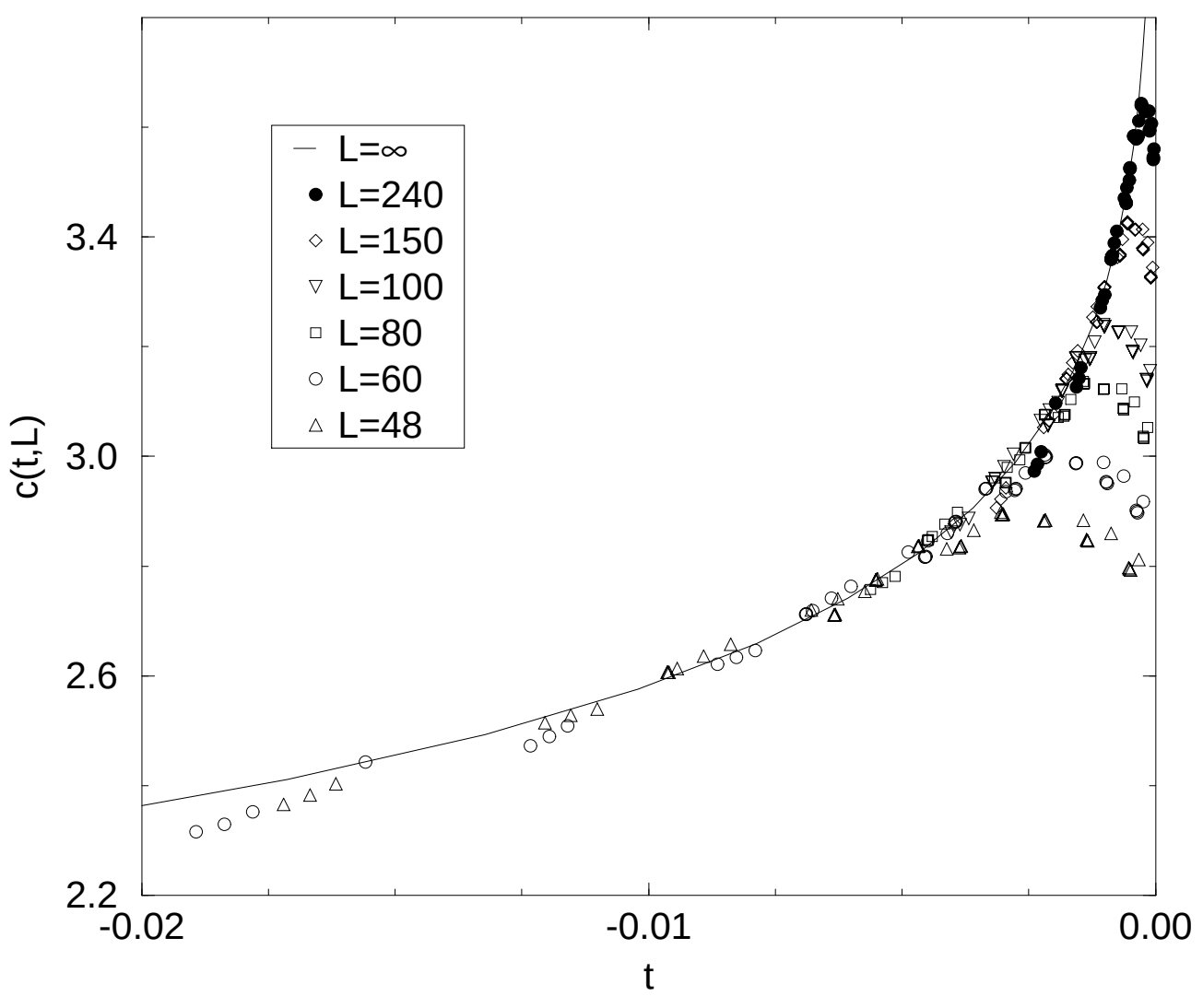

Figure 8: The specific heat for various lattice sizes $L$ derived from the scaling function $G_L(x)$. The solid line represents $L = \infty$ according to Eq. (21).

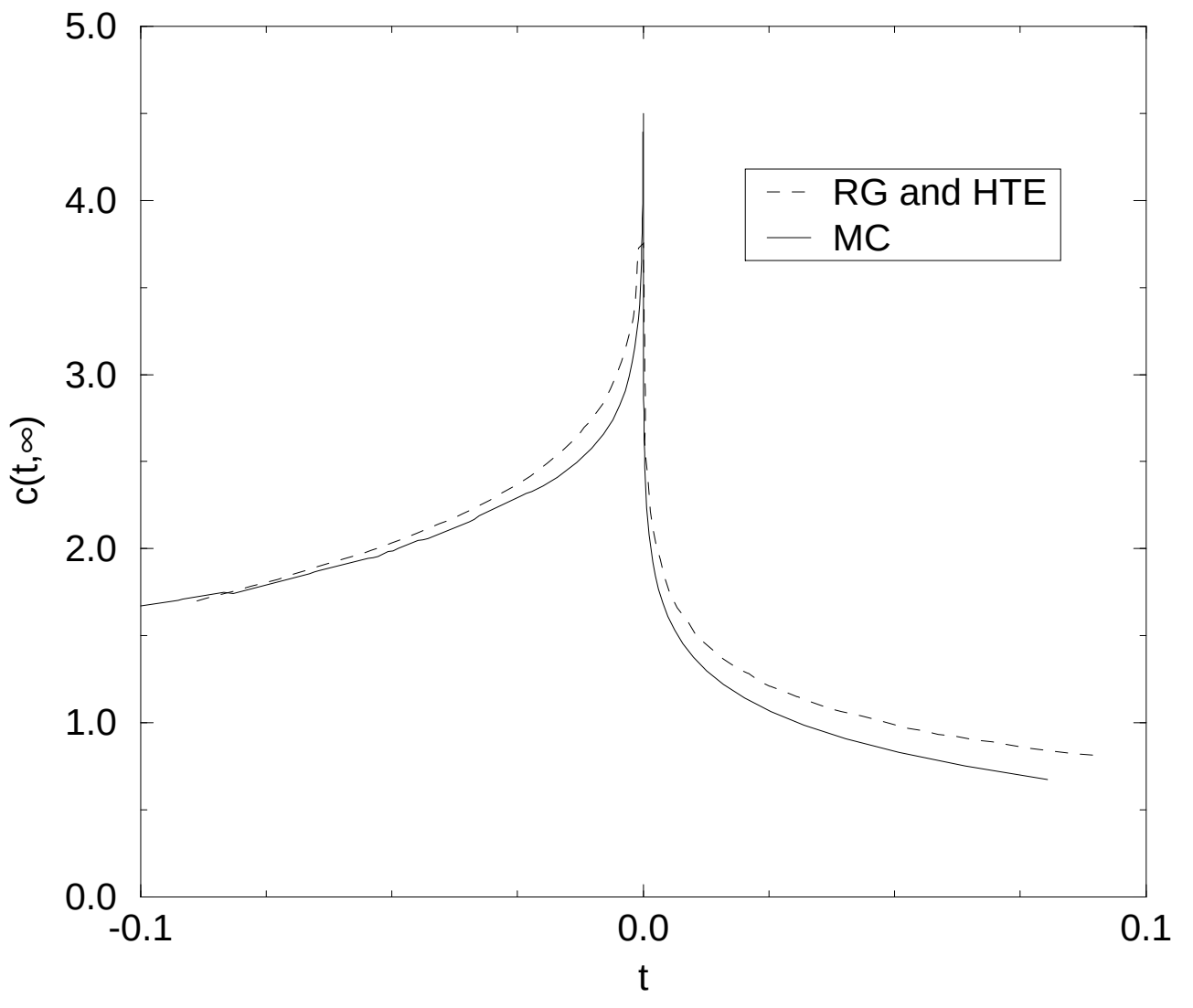

Figure 9: The bulk specific heat as a function of the reduced temperature. The solid lines represent the results of this work, the dashed lines represent the results of high temperature expansions and of Ref. [26].

In what follows we will determine the temperature dependence of the bulk specific heat above and below $T_\lambda$ using the experimentally determined values of $\nu = 0.6705$ [24] and $\alpha = -0.0115$ and thus the values for the quantities $c(0,\infty)$ and $c_1$ given by Eqs. (31) and (32). The knowledge of $g_\infty^+$ enables us to find the bulk behavior of the specific heat above $T_\lambda$ as given by Eq. (21) where

$$\tilde{c}_1^+ = c_1 g_\infty^+ = -30.5 \pm 1.1. \tag{35}$$

We found that below $T_\lambda$ the asymptotic form (21) is only valid for a narrow region near $T_\lambda$ (namely $|t| \ll 0.02$). In Fig.8 we show curves for the specific heat for lattice sizes ranging between $L = 48$ and $L = 240$. We computed these curves from our scaling function $G_L(x)$ where $\alpha = -0.0115$. The solid line in Fig.8 is the bulk curve. We identified the bulk values of the specific heat by the values of the specific heat for large enough size lattices which collapse onto one curve, because at those temperatures the specific heat does not feel the finite size of such large systems (because the correlation length for those temperatures is much smaller than the size of these lattices). Using larger and larger lattice sizes we can reach temperatures closer and closer to $T_\lambda$. Very close to $T_\lambda$ the bulk curve can be expressed by Eq. (21) giving

$$\tilde{c}_1^- = -29.225 \pm 0.025, \tag{36}$$

where $c(0,\infty)$ is taken from Eq. (31). In Fig.9 the bulk behavior of the specific heat is shown. The solid lines represent the results of the Monte Carlo calculation reported in this work. The dashed lines are taken from Ref. [26] which are a combination of the high–temperature series expansion of Ferer et al. (cf. Refs. [14]) and the renormalization group calculations of Ref. [26]. We would like to emphasize that the expression (21) is only valid in the interval $-0.02 \le t \le 0$.

We can use the universal ratio [27, 28]

$$\frac{\tilde{c}_1^+ \left(\xi_0^+\right)^3}{\tilde{c}_1^- \left(\xi_0^-\right)^3} \tag{37}$$

to compare our results to experimental results. The experimental value of $\xi_0^-$ is $\xi_0^- = 3.57$Å where $\xi_0^-$ is defined from $T/\Upsilon(t) = \xi_0^- |t|^{-\nu}$ according to Refs. [28, 29] ($\Upsilon$ denotes the helicity modulus). Using the $x-y$ model and the method of Ref.[23] we have calculated the helicity modulus for cubic lattices up to $40^3$ and using finite-size scaling for $\Upsilon$ and the above definition of $\xi_0^-$ we find $\xi_0^- = 1.21$ in lattice spacing units and thus $a = 2.95$Å. With $\xi_0^+ = 0.498a$ from Ref. [16] we find $\xi_0^+ = 1.47$Å ($\xi_0^+$ cannot be measured directly in the experiments). Since now the ratio $\xi_0^+/\xi_0^-$ is the same on the lattice and in the experiments so has to be the ratio $\tilde{c}_1^+/\tilde{c}_1^-$. We have

$$\frac{\tilde{c}_1^+}{\tilde{c}_1^-} = \begin{cases} 1.058(4) & \text{from Ref. [30]}, \\ 1.044(38) & \text{from this work.} \end{cases} \tag{38}$$

The agreement is quite satisfactory though it would be desirable to determine this ratio more accurately on the lattice, which requires larger lattices and probably longer simulation times as were used in the calculations reported here. We can also compare $A^\pm = \alpha \tilde{c}_1^\pm$ with the experimental results of Ref. [31] by expressing the specific heat of the $x-y$ model in physical units. This is accomplished by the equation

$$c_s = \frac{V_m k_B}{a^3} c = 14.74 \frac{\text{Joule}}{{}^\circ\text{K mole}} c, \tag{39}$$

where $V_m$ is the molar volume of $^4He$ at saturated vapor pressure at $T_\lambda$, $c$ is given by expression (5), and $c_s$ has units of Joule$({}^\circ$K mole$)^{-1}$. Using $\alpha = -0.0115$ and the values for $\tilde{c}_1^\pm$ given by Eqs. (35) and (36) we obtain

$$A^+ = 5.2 \pm 0.02, \tag{40}$$
$$A^- = 4.954 \pm 0.004, \tag{41}$$

with $A^{\pm}$ in units of Joule(°K mole)$^{-1}$. The experimental results of Ref. [31] are

$$\alpha = -0.009, \quad (42)$$
$$A^{+} = 5.82, \quad (43)$$
$$A^{-} = 5.504. \quad (44)$$

Also here the agreement is acceptable.

# 5 The film geometry

The specific heat on $L^2 \times H$ lattices in the limit $L \to \infty$ should obey:

$$\frac{c(t,H) - c(0,\infty)}{c(0,H) - c(0,\infty)} = G_H(tH^{1/\nu}). \quad (45)$$

The value of $c(0,\infty)$ is given either by (29) or (31) depending on the value for $\nu$.

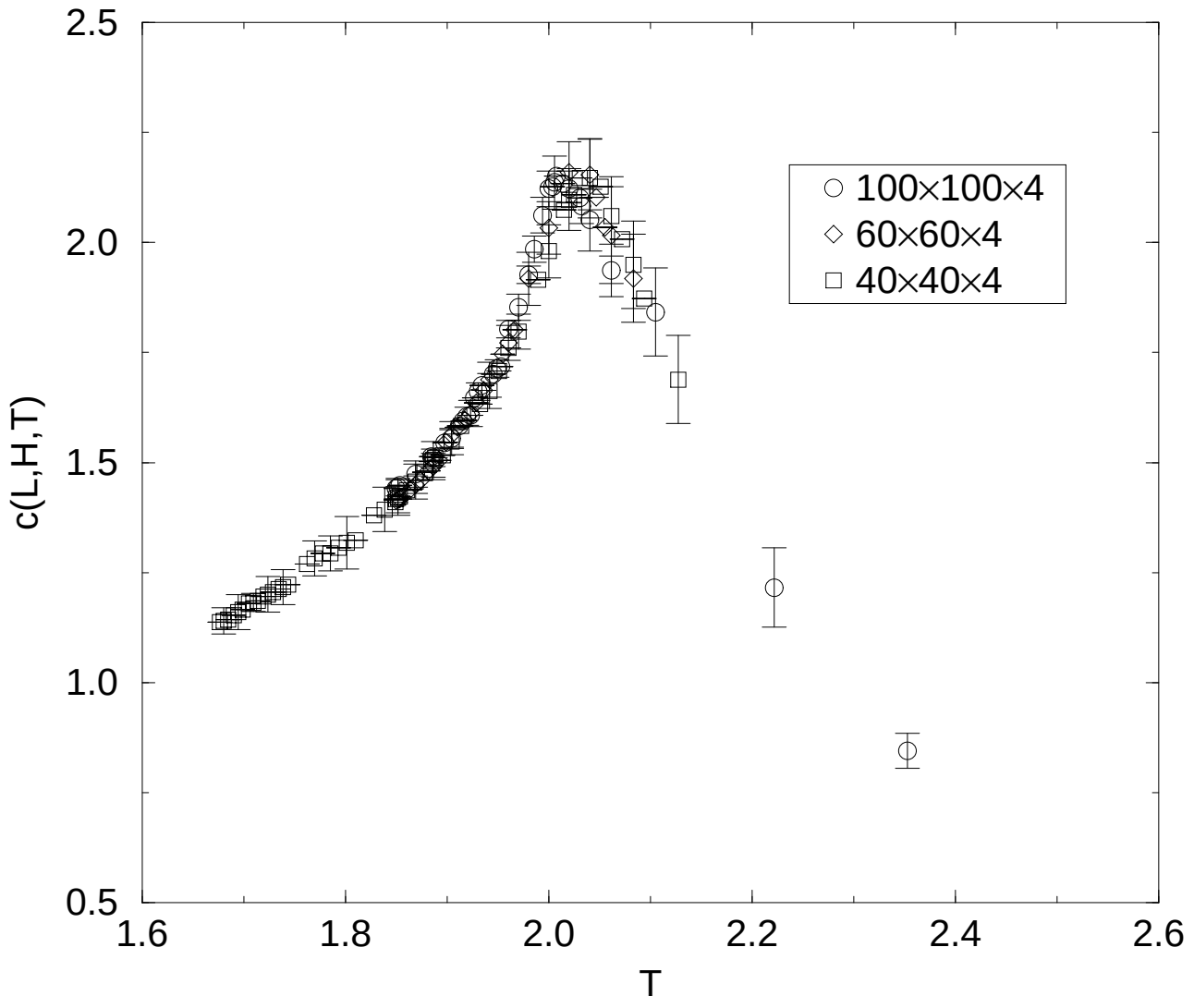

Figure 10: The specific heat $c(L,H,T)$ as a function of T for a film geometry for $H = 4$ and different values of $L$.

In order to apply the scaling form (45) we need to take the limit $L \to \infty$. However we find that the $L$–dependence of the specific heat is very weak, already for $L = 60$ and $L = 100$ for a fixed $H$ the specific heat data agree within error bars, as demonstrated in Fig.10. This weak $L$–dependence of the specific heat is quite in contrast to the very strong $L$–dependence of the helicity modulus, whose values had to be computed in the limit $L \to \infty$ at temperatures close to the $H$–dependent critical temperature $T_c^{2D}$ [23]. We take the values of the specific heat computed on $100^2 \times H$–lattices for $H = 6, 8, 10$ and assume those values to represent the case of infinite planar dimension.

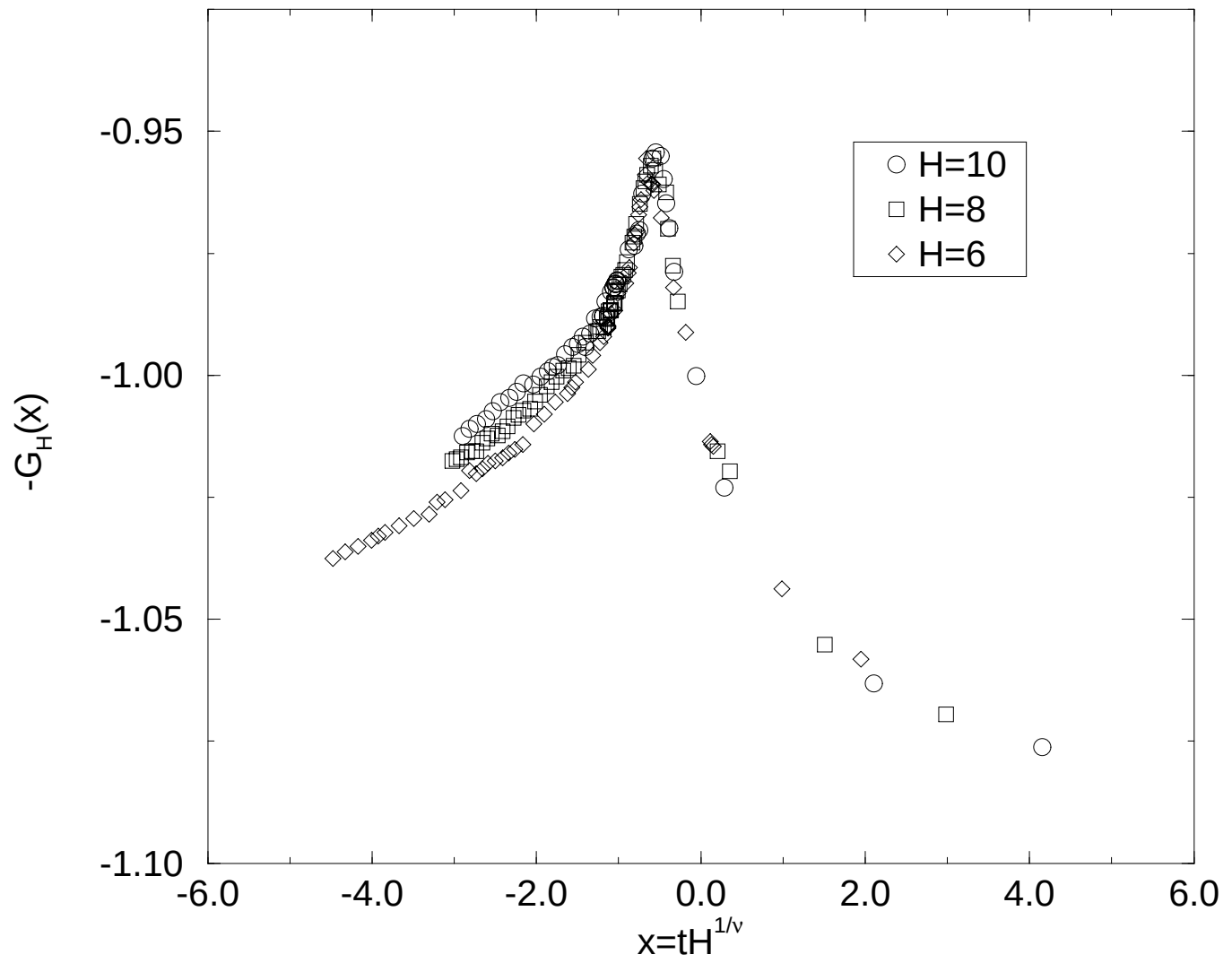

Figure 11: The scaling function $-G_H(x)$ (cf. Eq. (45)) for the film geometry. $c(0,\infty) = 20.45$, $\alpha/\nu = -0.0258$, $1/\nu = 1.487$, $T_\lambda = 2.2017$. The error bars have been omitted for clarity.

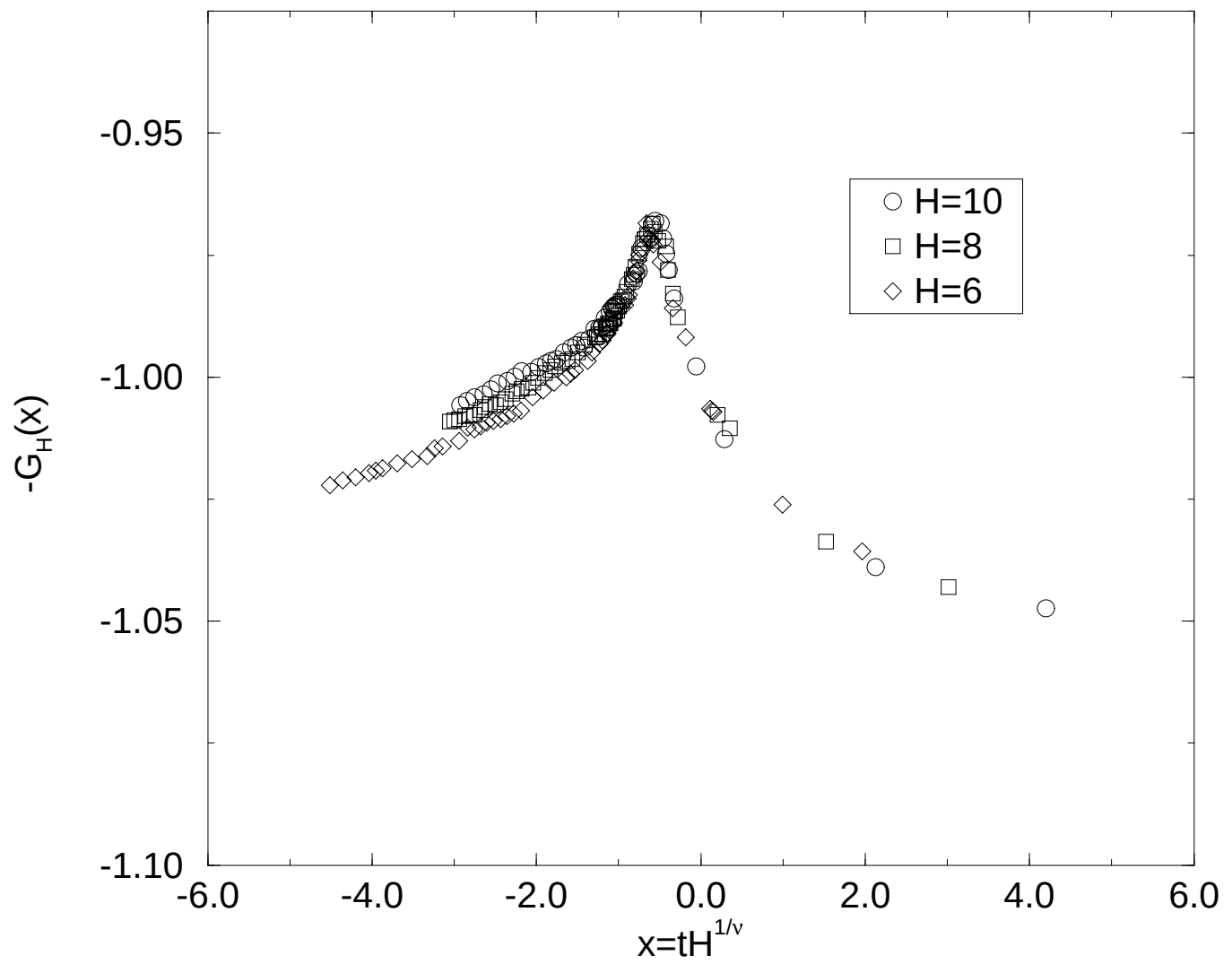

Figure 12: The scaling function $-G_H(x)$ (cf. Eq. (45)) for the film geometry. $c(0,\infty) = 30.3$, $\alpha/\nu = -0.0172$, $\nu = 0.6705$, $T_\lambda = 2.2017$. The error bars have been omitted for clarity.

In order to verify the expression (45), we plot $-G_H(x)$ in Fig.11 where the values for the parameters $\alpha$, $\nu$, $c(0,\infty)$, and $c_1$ are taken from Eqs. (27), (28), (29) and (30). The data collapse onto one universal curve. Instead if we use the experimentally found critical exponents $\nu = 0.6705$ and $\alpha = -0.0115$ and the values $c(0,\infty)$ and $c_1$ given by (31) and (32), respectively, to check the validity of scaling we obtain also data collapse as shown in Fig.12. The data points deviate from the scaling curve for $x \leq -1.5$ because we reach temperatures in this region which are outside the critical region. If we included data from much thicker films this deviation would occur for smaller values of $x$.

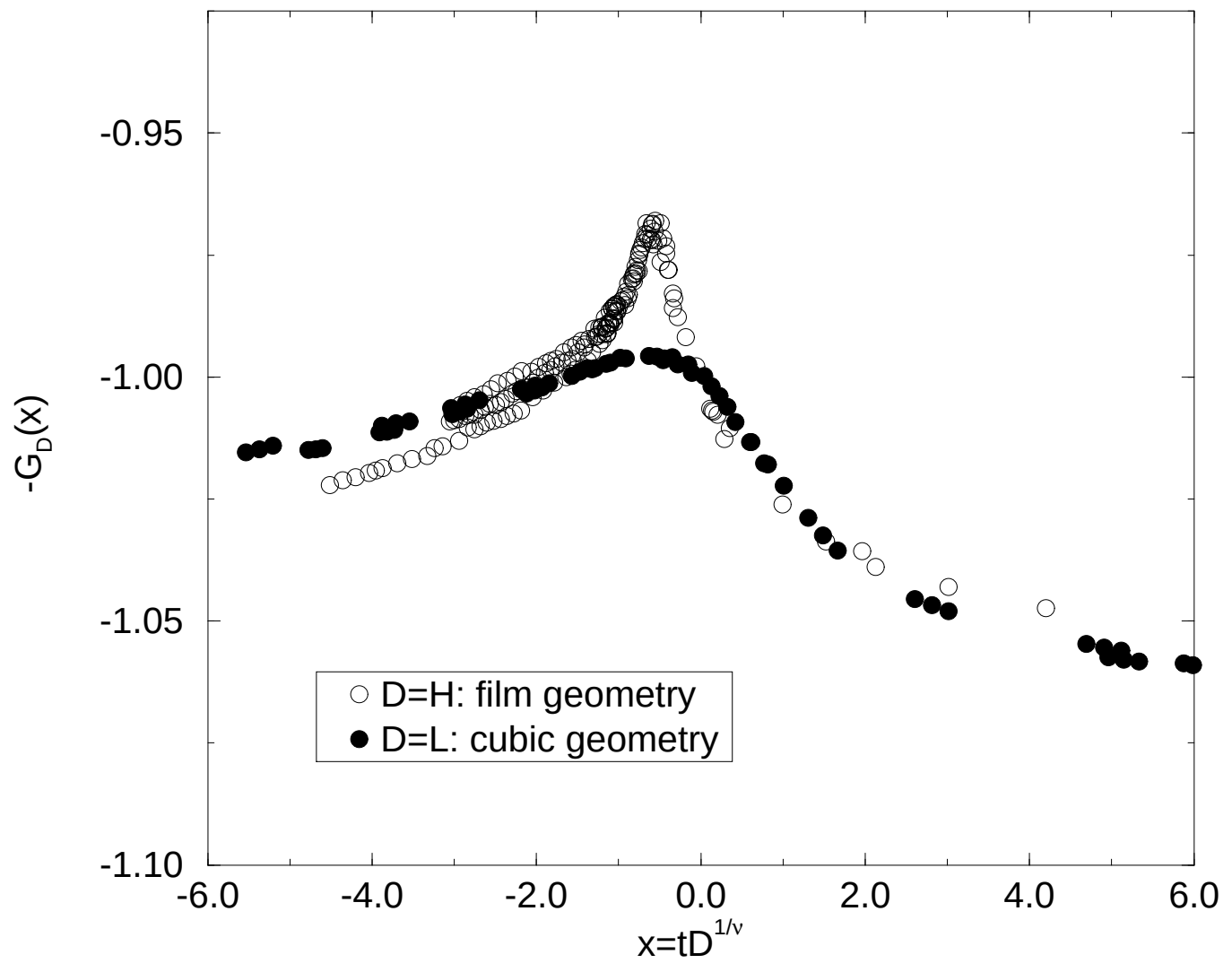

Figure 13: $-G_H(x)$ for the film geometry (circles) and $-G_L(x)$ for the cubic geometry (filled circles). $\alpha/\nu = -0.0115$, $\nu = 0.6705$, and $c(0,\infty) = 30.3$.

Fig.13 compares the universal functions $G_L(x)$ for the cubic geometry and $G_H(x)$ for the film geometry with $\nu = 0.6705$ and $\alpha = -0.0115$. The universal function $-G_H(x)$ has a sharper peak at its maximum.

Finally in this section we would like to compute the function $f_1(x)$ in physical units for the film geometry in order to make a direct comparison with experiments. We determined the lattice spacing in physical units $a$ and the value for $g_\infty^+$ in the previous section ($g_\infty^+$ is independent of the geometry and the boundary conditions). Thus, the function $f_1(x)$ in physical units (open circles) is determined up to a constant prefactor in front of the argument $x$ which is fixed by requiring that the maximum of $f_1(x_m)$ occurs at the same value of the argument $x_m$ as the maximum of the experimentally determined function $f_1(x_m)$. This constant prefactor turns out to be 1 and $f_1(x)$ is displayed in Fig.14. Also in this figure we plotted the experimentally determined function $f_1(x)$ of Refs. [9] (filled circles). The two functions are of the same order of magnitude but do not agree (varying the value of the constant prefactor does not improve the agreement). This is not surprising because the periodic boundary conditions only approximate the true physical boundary conditions. Therefore we also plotted the function $f_1(x)$ (solid line) obtained from a renormalization group calculation for the Landau–Ginzburg functional in a film geometry with Dirichlet boundary conditions [5, 6]. The importance of the correct physical boundary conditions is clear from this discussion.

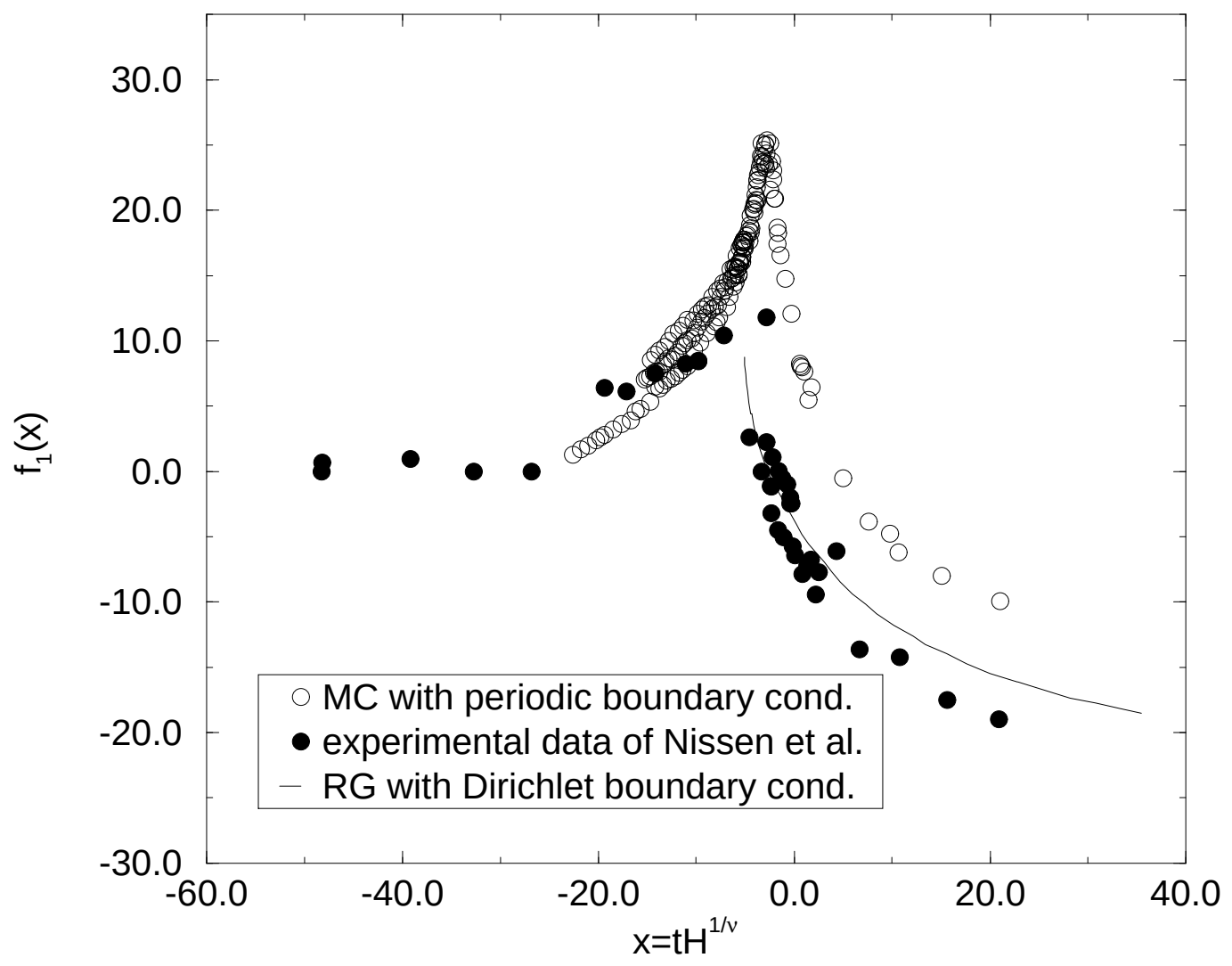

Figure 14: Comparison of the experimentally determined function $f_1(x)$ (filled circles) of Refs. [9], $f_1(x)$ obtained for films with periodic boundary conditions (open circles), and $f_1(x)$ for films with Dirichlet boundary conditions of Ref. [6]. $f_1(x)$ has the units Joule($^\circ$K mole)$^{-1}$ and $H$ is in Å.

# 6 Summary

We have investigated the finite–size scaling properties of the specific heat of the $x-y$ model in a cubic geometry $L \times L \times L$ and in a film geometry $L \times L \times H$ with $L \gg H$. Periodic boundary conditions were applied in all directions. For the cubic geometry we find strong evidence that the critical exponent $\alpha$ is negative, so the specific heat does not diverge at the critical temperature, in qualitative agreement with experimental findings. However, we were not able to determine $\alpha$ very accurately, we find $\alpha/\nu = -0.026(8)$, $1/\nu = 1.49(8)$, which is in reasonable agreement with experiments. Our values for $\alpha$ and $\nu$ fulfill the hyperscaling relation. We confirmed the scaling assumptions (17) for the cubic case and (45) for the film geometry. We also used the experimentally determined values for $\nu$ and $\alpha$ to compute the scaling functions for the specific heat in the two geometries. We derived the bulk behavior of the specific heat below and above the critical temperature and compared these results to renormalization group calculations and high–temperature expansions. Good agreement between the scaling function $f_1(x)$ (cf. Eq. (18)) for the cubic geometry obtained from our Monte Carlo simulation and renormalization group calculations was found. For the film geometry we compared $f_1(x)$ derived from our Monte Carlo data to the experimentally determined function $f_1(x)$ and the renormalization group result of Refs. [5, 6]. This comparison leads to the conclusion that the boundary conditions determine the shape of the universal functions and have to be chosen properly in order to find agreement with the experimental results.

## 7 Acknowledgements

This work was supported by the National Aeronautics and Space Administration under grant no. NAGW-3326.